# Use Dimensionality Reduction and SVM Methods to Increase the Penetration Rate of Computer Networks

**Authors**: Amir Moradibaad [1,*], Ramin Jalilian Mashhoud [2]

1. Polytechnic University of Milan, Milan, Italy     [amir.moradibaad@mail.polimi.it](amir.moradibaad@mail.polimi.it)
2. Polytechnic University of Milan, Milan, Italy     ramin.jalilian@mail.polimi.it

**Abstract**

In the world today computer networks have a very important position and most of the urban and national infrastructure as well as organizations are managed by computer networks, therefore, the security of these systems against the planned attacks is of great importance. Therefore, researchers have been trying to find these vulnerabilities so that after identifying ways to penetrate the system, they will provide system protection through preventive or countermeasures. SVM is considered as one of the major algorithms for intrusion detection. One of the major problems is the time of training and the need to improve its efficiency when it comes to work with large dimensions. In this research, we try to study a variety of malware and methods of intrusion detection, provide an efficient method for detecting attacks and utilizing dimension reduction. Thus, we will be able to detect attacks by carefully combining these two algorithms and pre-processes that are performed before the two on the input data. The main question raised in this study is how we can identify attacks on computer networks with the above-mentioned method. In anomalies diagnostic method, by identifying behavior as a normal behavior for the user, the host, or the whole system, any deviation from this behavior is considered as an abnormal behavior, which can be a potential occurrence of an attack. In this research, the network intrusion detection system is used by anomaly detection method that uses the SVM algorithm for classification and SVD to reduce the size. The various steps of the proposed method include pre-processing of the data set, feature selection, support vector machine, and evaluation. The NSL-KDD data set has been used to teach and test the proposed model. In this study, we inferred the intrusion detection using the SVM algorithm for classification and SVD for diminishing dimensions with no classification algorithm. And also the KNN algorithm has been compared in situations with and without diminishing dimensions and the results have shown that the proposed method has a better performance than comparable methods.

## 1- Introduction

Most of the planned attacks abuse software fault and security gaps and through malware called bad bad-ware. Generally, any type of software code that runs on a computer system and performs an unwanted operation is known as malware. Since it's impossible to completely remove software errors, all software has security holes, which refers to software vulnerabilities. Therefore, researchers have been trying to find these vulnerabilities so that after identifying ways to penetrate the system, they will provide system protection through preventive or countermeasures. The most common types of attacks toward Cyber-Physical systems are against energy infrastructures and specifically power system grids. When the attacker injects the false data into measurement devices and changes the system state variables including electricity price for retailers or end-users [1]. Support vector machines (SVMs) have been recognized as one of the most successful classification methods [2-5]. The learning ability and computational complexity of training in support vector machines may be independent of the dimension of the feature space; reducing computational complexity is an essential issue to efficiently handle a large number of terms in practical applications of text classification [6]. SVM is considered as one of the key algorithms for intrusion detection. One of the major problems is the time of training and the need to improve its efficiency in work with large dimensions. A novel methodology that leverages-optimization [7] to detect sparsity is compressive sensing which has emerging applications including dimension reduction [8]. In this research, we try to study types of malware and methods of intrusion detection, provide an efficient method to explore the attack and use the Dimension Reduction method. Thus, we will be able to detect attacks by carefully combining these two algorithms and pre-processes that are performed before the two on the input data. The main question raised in this study is how we can identify attacks on computer networks with the above-mentioned method [9]. Today, most of the vital infrastructure of a city and a country is controlled and managed by computer networks; therefore, security of intrusion detection systems against scheduled attacks is of great importance. These types of systems are essential for many organizations, from small offices to large multinational corporations. One of the benefits of these types of systems is the increased efficiency of intrusion detection compared to manual systems, having a source of full knowledge of attacks, the ability to handle a large amount of information [10], the possibility of a warning that will reduce the relevant damage, increased deterrence and the ability to report appropriately. In fact, without having an intrusion detection system in the computer networks, these networks will sooner or later be fired by various attacks, causing extensive and irreparable damage to urban and national affairs [11]. In reference [12] various ways of performing dimensionality reduction on high-dimensional microarray data summarized. In [13] the authors used an innovative method to radically reduce the size of the data and to select the useful features. Many different feature selection and feature extraction methods exist and they are being widely used. All these methods aim to remove redundant and irrelevant features so that classification of new instances will be more accurate. Analyzing microarrays can be difficult due to the size of the data they provide. In addition the complicated relations among the different genes make analysis more difficult and removing excess features can improve the quality of the results[14]. We present some of the most popular methods for selecting significant features and provide a comparison between them. Their advantages and disadvantages are outlined in order to provide a clearer idea of when to use each one of them for saving computational time and resources.

Reference [15] introduces two new methods of dimension reduction to conduct small-sample size and high-dimensional data processing and modeling [16]. Through combining the support vector machine (SVM) and recursive feature elimination (RFE), SVM-RFE algorithm is proposed to select features, and further, to add the higher order singular value decomposition (HOSVD) to the feature extraction which involves successfully organizing the data into high order tensor pattern. The validation of simulation experiment data shows that the proposed novel feature selection and feature extraction methods can be effectively applied to the research work for analyzing and modeling the data of atmospheric corrosion. The feature selection method pledges that the remaining feature subset is optimal; feature extraction method reserves the original structure, discriminate information, and the integrity of data, etc. Finally, they propose a complete data dimensionality reduction solution that can effectively solve the high-dimensional small sample data problem, and code programming for this solution has been implemented.

Kumar et al (2017) discusses the approach for intrusion detection and classification by devising a membership function, inspired from Yung, Jung, & Shie-Jue (2014) and used in this work to carry the dimensionality reduction of processes present in the training set in evolutionary approach. The reduced process representation may then be used to perform classification and prediction for detecting intrusion. It is seen that the reduced representation of processes retains the system call distribution of the initial process. Experiment results show the proposed approach is better compared to existing approaches and helps in effective identification of U2R and R2L attacks [17].

Discriminant function is very critical in separating the normal and anomaly behavior accurately. The support vector machine based classification algorithm is used to classify the intrusions accurately by using the discriminant function. The effective discriminant function will be accurately identifies the data into intrusion and anomaly. The evaluation of the discriminant is important in the evaluation of the intrusion detection system. Performance of intrusion detection system depends on the choice of the discriminant function [18].

Methods based on k-NN are hybrid methods presented [19,20]. They have provided five methods: DROP5, .., DROPI; these methods are based on the concept of dependency. Template affiliations p are patterns where p is one of k closest neighbors. DROPI deletes the pattern p from T if p's dependencies are properly classified in S without p; by virtue of this rule, since the affiliations of a pattern of noise without that pattern can be properly categorized, DROPI eliminates the noise patterns. But in DROPI, when the neighbors first remove a noise pattern, then the noise pattern will not be deleted. In order to solve this problem in DROP2, affiliates of a template are searched for in the entire training suite. That is, p is only deleted when its affiliations are classified in T without p. DROP3 and DROP4 initially remove the noise patterns with a ENN-like filter, and then apply DROP2. DROP5 operates on the basis of DROP2, with the difference that in order to smooth the decision boundary, first, it removes the closest rivals (the closest patterns to the different classes).

Another category of research is based on the direct calculation of the decision level, which is based on the separation margin of the maximum separation of data from two classes and is one of the group of methods of separation. In this category of research, we try to obtain a decision level for the classification of data. A level of decision able to create the maximum margin of separation between the two classes. The two most famous methods in this area are the works [21-23]. In [21], the goal is to determine the separation function based on distance with the maximum distance from the data of each class in Banach space. For this purpose, the Liebix function is used in which a Liebixitz function constant denotes the margin of the separator. Finding the appropriate Libsitz function for real data is a problem and the current solutions are inefficient in terms of performance. Its implementation results are not satisfactory in terms of performance. In another method, SVM [22] and [23], with the aim of considering the maximum separation margin between two classes of data, with the optimization procedure, a linear decision level is presented in Hilbert space. In this way, the problem of finding the decision level has become a second-order planning problem. Solving this planning problem is one of the current challenges for high-volume issues such as intrusion detection.

## 2.2 Dimensionality reduction methods

Because many researches including this research have been used to reduce the dimensions of intrusion detection, we have explored the methods of dimensionality reduction. The progress made in collecting data and storage capabilities over the past decades has led to a large amount of information in many sciences. Researchers in many fields, such as engineering, astrology, biology and economics, are faced with more and more observations each day. Compared to older and smaller databases, today's data bases have created new challenges in the area of data analysis [24].

Traditional statistical methods have lost their effectiveness for two reasons today. The first reason is to increase the number of observations, and the second reason, which is more important, is to increase the number of variables associated with observation. The number of variables to be measured for each observation is called the dimension of data. The "variable" is used more often in statistics while in computer science and machine learning more than the terms "feature" or "attribute" are used.

Methods for data dimensionality reduction are divided into two categories:

• Methods based on feature extraction: These methods map a multidimensional space into a smaller space. In fact, by combining the values of existing attributes, fewer features, such that these features have all (or a large part) of the information contained in the original features [25]. These methods are divided into linear and nonlinear groups[26].

• Methods based on feature selection: These methods attempt to reduce the dimensions of the data by selecting a subset of the initial properties [27]. At times, data analyzes such as classification on a reduced space are better than the original space.

## 3. Research methodology

Empirically, through the study of papers and scientific research, technical reports, books, dissertations and research projects, the material contained in scientific and technical resources in the field of SVM as well as techniques for the detection and assessment of influence will be conducted and then in vitro, by simulating the proposed method using SVM and dimming dimension for intrusion detection. This research has several stages that we are reviewing here. In the preprocessing section, the inputs to the system are subjected to initial processing and the values of the attributes are normalized. In order to make the calculations more precise and reduce the computational time, we will reduce the number of dimensions of the input data by dimming the dimension to the features that are additional and are not needed in the calculations, and only the features of the input data that are effective in computing are effective. Now, using educational data, we will train the system and thus the system will learn the pattern of attacks in terms of input characteristics. In this way, the model is also formed, which is the basis of the forecast. In the next step, using the learning model we created in the previous section, we can perform intrusion detection into the system in terms of input data and determine what, in what cases, penetration will occur. Finally, with regard to the system output in detecting network penetration, we evaluate how accurately the intrusion detection is done by the system, and we obtain precision in the form of precision. Our purpose of the proposed method is to provide a classification-based method for intrusion detection that improves SVM learning time and precision. Based on the combination of the proposed method, the ability to identify types of influences increased. Features of the proposed method include combining dimension reduction with SVM for classification and also the use of pre-processing data. In this chapter, we first discussed the process of the proposed method then, each step of the proposed method is described.

### 3-1. Process of proposed method

The following figure shows the process of this research.

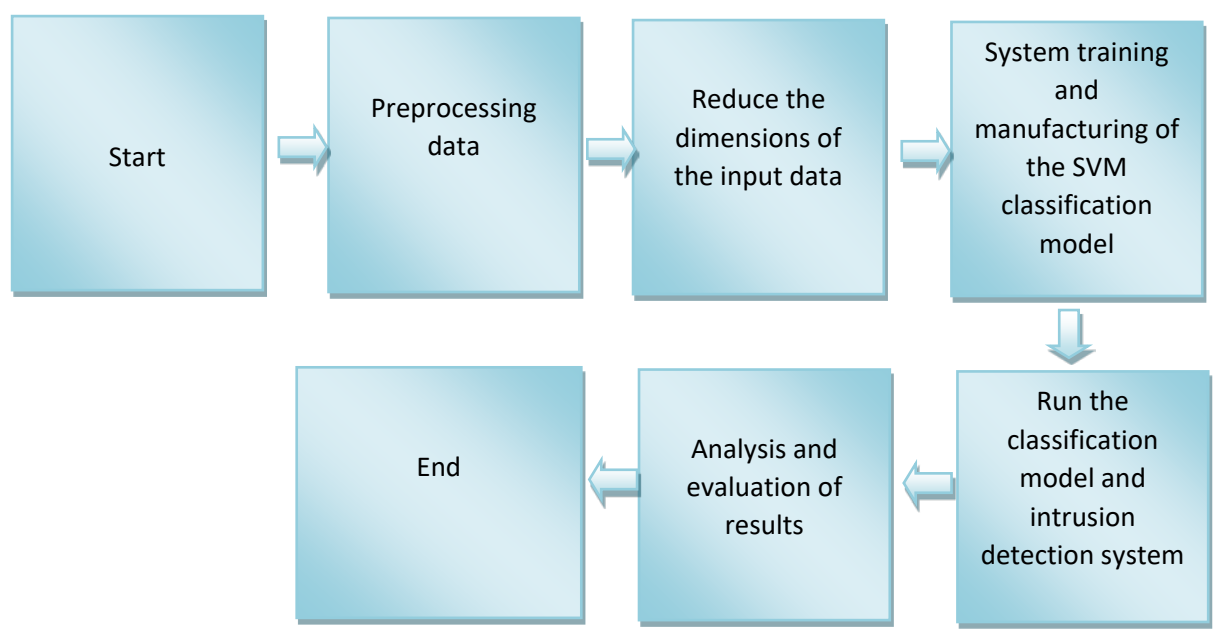

Figure 1. The process of conducting research

In the preprocessing section, the inputs to the system are subjected to initial processing and the values of the attributes are normalized. In order to make the calculations more precise and reduce the computational time, we will reduce the number of dimensions of the input data by dimming, so that features that are extra and not needed in the calculations are eliminated and only the characteristics of the input data remain effective in computing. Now, using educational data, we will train the system so that the system learns the pattern of attacks in terms of input characteristics. In this way, the model is also formed, which is the basis of the forecast. In the next step, using the learning model we created in the previous section, we can perform intrusion detection into the system in terms of input data and determine what, in what cases, penetration will occur. Finally, with regard to the system output in detecting network penetration, we evaluate the extent to which the intrusion detection is done correctly by the system and we get precision in form of precision.

**4.3 Data dimensionality reduction**

Here, we use the SVD method to reduce the dimensions of the data, and this method works based on the matrix decomposition. SVD can convert it to a small dimensional matrix in order to remove irrelevant information from a large-scale documentary word matrix. Assume that the matrix X, which is a matrix $t \times d$, is constructed so that t is the number of keywords and d is the number of data. Each element $X[t,d]$ shows a special feature. We define X as follows: we also have $X = USV^T$: so that the elements of S are all regular values of X.

We define $n = \min\{t, d\}$ and show single values with $\sigma_1 \geq \sigma_2 \geq ... \geq \sigma_n \geq 0$. U and V are respectively $d \times d$ and $t \times t$. After processing by SVD, the equation $X = USV^T$ is simplified $X_k = U_k S_k V_k^T$ as shown in the figure below. Dimensions $U_k$ $S_k$ and $V_k^T$ decreases to $d \times k$, $k \times k$ and $k \times t$. The value of the element k is smaller than the vector space of the original space. $S_k$, K holds a greater single value in the matrix. Also $U_k$ is a vertical vector and $V_k$ is a horizontal vector. SVD is used to decompose $X_i$ and obtain three matrices V, S, and U.

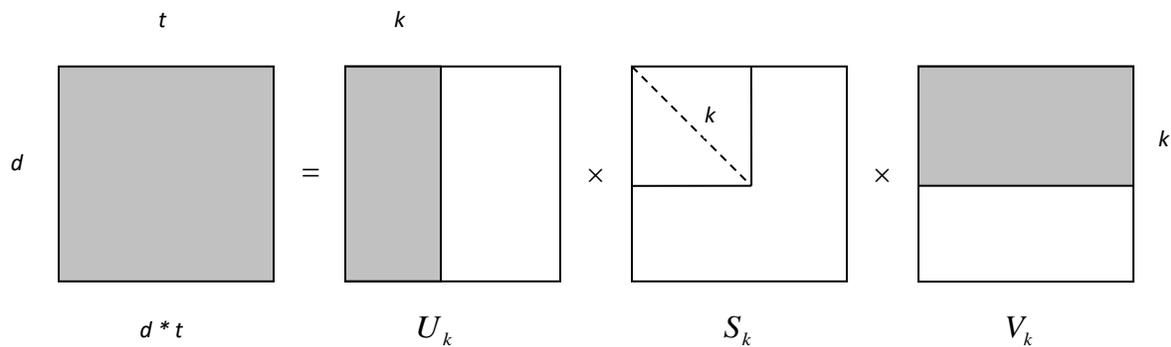

Figure 2 Using SVD for the decomposition of the $d * t$ matrix

This method works by factoring the important factors of the input data matrix. Assume that the matrix M is a mxn matrix. Then the following equation for the matrix M will be divided into three other matrices.

M = USV          (1)

The matrix U is a unit matrix $m_x m$, the matrix S is a diagonal matrix $m_x n$ whose diagonal lines are nonlinear and V is a unit $n_x n$ matrix.

The diagonal members of the S matrix, which are actually $S_{i,i}$, are known as singular values of the matrix M. m is column of the U and n matrix are the columns of the matrix V, respectively, as left-singular vectors and right-singular vectors of the matrix M.

To further explore the properties of this relationship, we need to introduce new concepts called Eigenvalues and Eigenvectors. Assuming that A is a square matrix and V is a non-dimensional vector and λ is a numerical value, then if the following equation holds, the vector V as the Eigenvector of the matrix A and the values λ will be the Eigenvalue of the matrix A.

A V = λV    (2)

### 5.3 Intrusion detection using SVM-based classification

Using SVM on classification issues is a new approach that has been considered in recent years and has been used in a wide range of applications. The SVM approach is that in the training phase, the boundary of decision making is chosen so that the minimum distance to each of the categories is maximized. This kind of choice makes our decision-making process in a good way to withstand the noise conditions and have a good response. This is how borders are selected based on points called backward vectors. Suppose we have a collection of data points and we want to divide them into two categories $c_i = \{-1, 1\}$. Each $x_i$ is dimensional vector $P$ of real numbers, which in fact are the same variables that represent the behavior of the software. Linear classification methods try to separate the data by constructing an abstract (linear equation). The support vector machine classification method, one of the linear categorization methods, finds the best astronomy that separates the data from the two classes with maximum distance. Support vector separators in the input space create linear boundaries. By extending the space of input to a larger space, using the base functions, such as polynomials, the degree of freedom of this process can be increased. In general, the linear boundaries in the new input (extended) field create a better separation in the classroom and turning them back into the initial input space into non-linear boundaries. Therefore, we can say that the idea of a vector machine consists of two processes:

- Non-linear mapping of input vector to a higher dimensional space that is hidden from the input and output view.
- Creating an optimized super-page to differentiate the features that have been achieved at the creation stage

### 2.3.5 Designing supporting vector machines for classification

For designing supporting vector machines, we can use (3) and solve it. This relationship can be represented in such a way as to include only the internal multiplication of the vector and the new features.

$$L_P = \tfrac{1}{2} \|w\|^2 + \gamma \sum_{i=1}^{n} \xi_i - \sum_{i=1}^{n} \alpha_i \left[ y_i (x_i^T w + w_0) - (1 - \xi_i) \right] - \sum_{i=1}^{n} \mu_i \xi_i \qquad (3)$$

With zero we have the corresponding derivatives:

$$w = \sum_{i=1}^{n} \alpha_i y_i x_i \qquad (4)$$

$$0 = \sum_{i=1}^{n} \alpha_i y_i, \qquad (5)$$

$$\alpha_i = \gamma - \mu_i, \quad \forall i \qquad (6)$$

Non-negative constraint: We for each *i* have $\alpha_i, \mu_i, \xi_i \geq 0$. Putting relations (4) to (6) in relation (7), the objective function of the Lagrangian Duality (Woolf) is obtained:

$$L_D = \sum_{i=1}^{n} \alpha_i - \frac{1}{2} \sum_{i=1}^{n} \sum_{j=1}^{n} \alpha_i \alpha_j y_i y_j x_i^T x_j \tag{7}$$

In this case, the Lagrangian Duality Function (7) is in the form of the equation (8).

$$L_D = \sum_{i=1}^{n} \alpha_i - \frac{1}{2} \sum_{i=1}^{n} \sum_{j=1}^{n} \alpha_i \alpha_j y_i y_j \langle \varphi(x_i), \varphi(x_j) \rangle \tag{8}$$

From (1) (6) we can write the answer function $f(x)$ in the form (9):

$$\begin{aligned} f(x) &= \varphi(x)^T w + w_0 \\ &= \sum_{i=1}^{n} \alpha_i y_i \langle \varphi(x), \varphi(x_i) \rangle + w_0 \end{aligned} \tag{9}$$

As before, by obtaining $\alpha_i$, we can obtain $w_0$ with the solution $f(x_i) = 0$ in (9) for each (or all) $x_i$ that is $0 < \alpha_i < \gamma$. According to the above, both relations (8) and (9) are conveyed by internal multiplication and not directly includes $\varphi(x)$. Therefore, there is no need to specify the transformation $\varphi(x)$, and we only need to have the relation of the core function, which is the multiplication of the interior in the transformed space.

$$K(x, x') = \langle \varphi(x), \varphi(x') \rangle = \varphi^T(x) . \varphi(x') = \sum_{j=0}^{M} \varphi_j(x) \varphi_j(x') \tag{10}$$

K must be symmetric and (semi) positive function. From (11), it is possible to writ the answer function as follow.

$$\hat{f}(x) = \sum_{i=1}^{n} \hat{\alpha}_i y_i K(x, x_i) + \hat{w}_0 \tag{11}$$

The following figure shows the support vector machine architecture. The three nuclear functions commonly used in support vector machines are the polynomial of degree d, the radial base function (RBF), and the sigmoid (perceptron) function, whose relationships are in (12), (13), and (14).

$$K(x, x') = (1 + \langle x, x' \rangle)^d \tag{12}$$

$$K(x, x') = \exp(-\|x - x'\|^2 / c) \tag{13}$$

$$K(x, x') = \tanh(\kappa_1 \langle x, x' \rangle + \kappa_2) \tag{14}$$

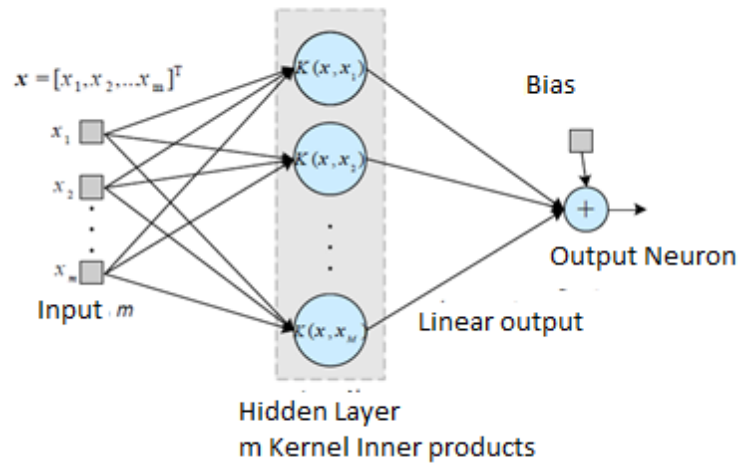

Figure 3- Support Vector Machine Architecture

The most serious problem in the SVM method is the selection of the kernel function. Several methods and principles have been introduced for this work: Diffussion kernel, Fisher kernel, String kernel, etc., and research is also under way to obtain a kernel matrix from existing data.

The following figure shows the performance of supporting vector machines with different cores in the classification of two-class data with a two-dimensional Gaussian probability density function with the same variance and averages. Only training data is shown in the form. The percentage error of classification of training and test data is presented in Figures.

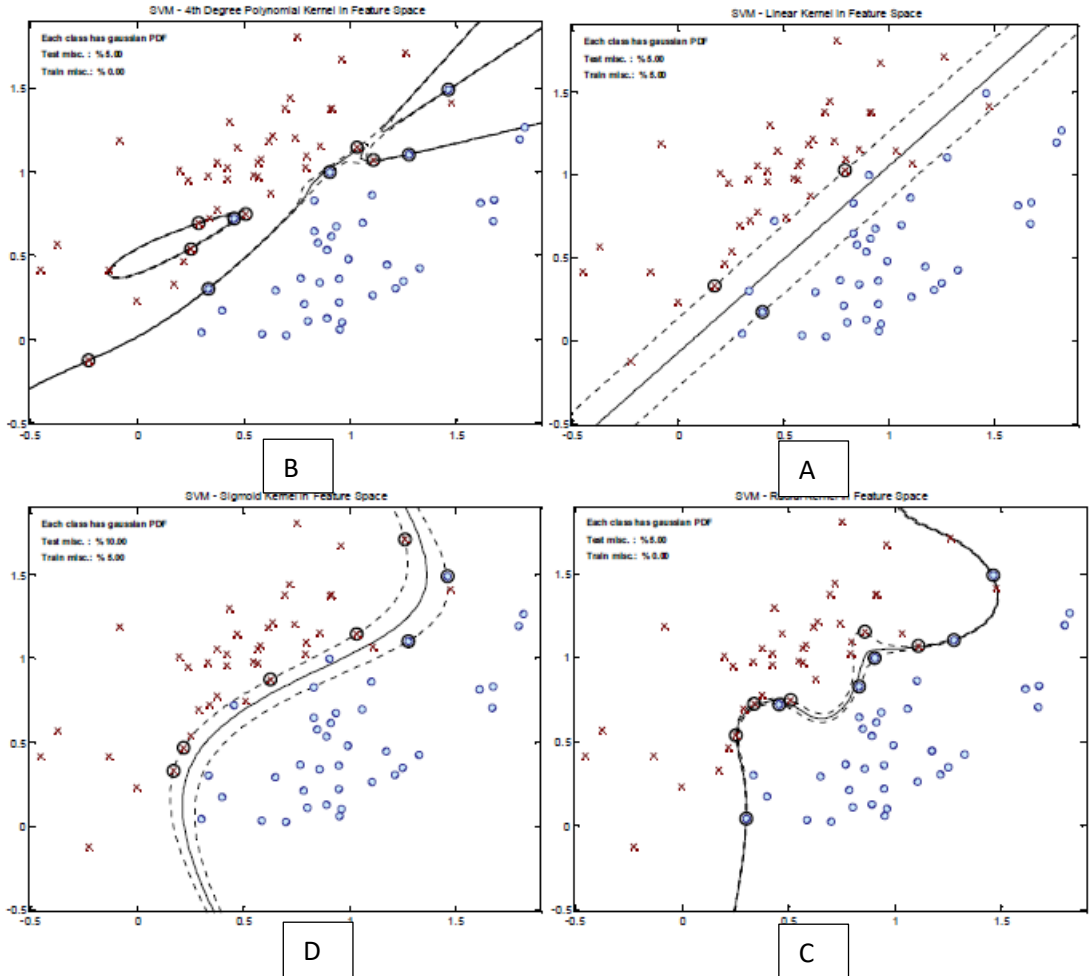

Figure 4. Support vector machine with a) Linear b) Fourth order polynomial c) Radial d) Sigma to isolate data with Gaussian distribution

In practice, a polynomial kernel or radionuclide core function (RBF kernel) with good width is a good start. Note that SVM with RBF core is very close to RBF neural networks, with radial base functions that are automatically selected for SVM. For this purpose, the RBF core of Equation 15, which is most used in this case, has been used.

$$K(X_i, X_j) = e^{-\|X_i - X_j\|^2 / 2\sigma^2} \qquad (15)$$

## 4. Analysis and evaluation of results

### 4.1 Data collection used in this research

To implement, measure and evaluate the proposed method of this study, the NSL-KDD dataset [28] has been used, which includes selected records from KDD-CUP99 [29]. And the problems in that dataset, as well as duplicate records, have been resolved. The KDD-CUP99 dataset since 1999 has been considered as one of the most commonly used datasets to evaluate malware detection and network penetration techniques. This dataset was developed by Stollfo et al. [30] based on the DARPA'98 Intrusion Detection Evaluation System [31] which is a collection of about 5 million records of various connectivity, each of which is about 100 bytes in size. The KDD training suite consists of a total of 4900000 vectors of different connections, each vector consisting of 41 features.

Table 1. Approximate distribution of training and experimental data in the NSL-KDD dataset

| Class | Training | Experimental |
|---|---|---|
| NORMAL | 48% | 19% |
| PRB | 20% | 1% |
| DOS | 26% | 73% |
| U2R | 0.2% | 0.07% |
| R2L | 5% | 5% |

**4-2 SVM modeling for intrusion detection**

The SVM-based model for intrusion detection is shown in the figure below. In this process model, data is called two paths, one of which is the system training data and the other test data. By reading the training data, pre-processing is performed first on the data, and data is prepared for the next steps. Then, the SVM model is based on training data. The second part of the data that is read and test data that they also after preprocessing, to evaluate and calculate precision, they are used. You can refer to the RapidMiner site to get familiar with the operators used in this process.

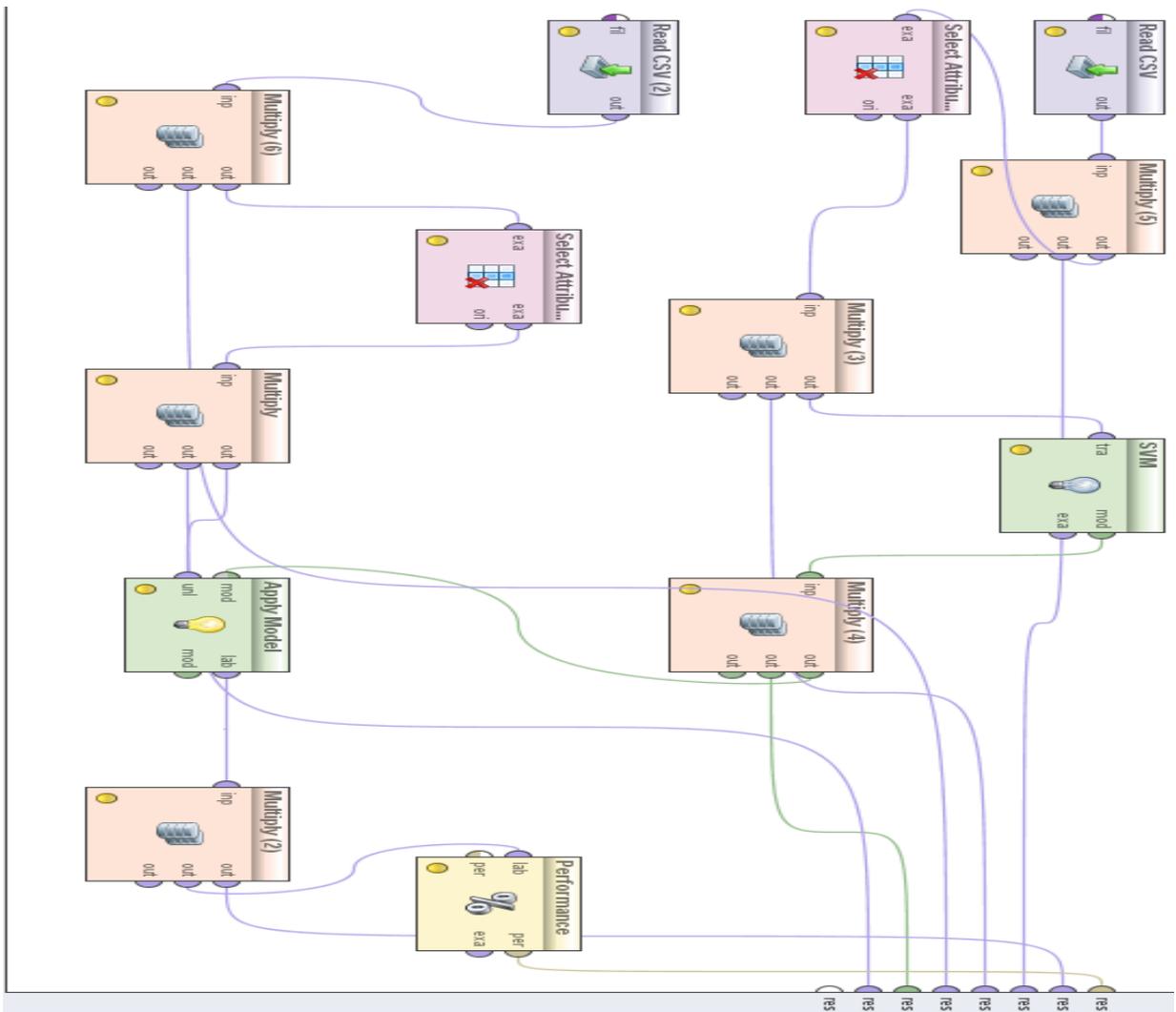

Figure 5 Intrusion detection modeling in the Rapiminer environment

**4-3 Evaluation of the results**

The purpose of the proposed method of intrusion detection system is to achieve improvement in intrusion detection efficiency, so that the best response should be considered with regard to the type of intrusion. In fact, if the classification of the attacks is properly determined, the choice of the right decision for each type of attack is taken according to the case and according to the decisions of the network leader. Of course, these decisions can be stored as a response to any kind of attack on a series of responses to the security system so that it can be possible to detect any impact of the security measures necessary to prevent the system from being compromised. These arrangements can depend on the type of network, organizational policies, and other various factors. The goodness of the results should be calculated on the basis of metrics, so that it can be quantitatively measured. Therefore, the two metrics that are considered in this section are precision and recall, which are well-known and widely used in the evaluation of data mining or machine learning algorithms. The exact definition of precision is the following:

$$Precision = \frac{TP}{TP + FP}$$

And the definition of recall is also as follows:

$$Recall = \frac{TP}{TP + FN}$$

In the above formulas, TP represents the number of data that is correctly assigned to the positive class, FP represents the number of data that is misidentified to the positive class, and FN represents the number of data that is mistakes are assigned to the negative class. It should be noted that for the purpose of the positive and negative class, there are two cases considered for the given data in a classification problem.

$$F - Measure = \frac{2 \times Precision \times Recall}{Precision + Recall}$$

Given the criteria mentioned in this section, the results can now be presented. To train the proposed system, 23,000 records from the KDDTrain + series were randomly selected which includes 24 known attack types, and 5,500 records from the KDDTest + series were selected to test, in addition to 24 known attacks, including 14 unknown new attacks. After selecting features, 9 features were deleted.

Comparison of the precision, recall and F-score by separating the attack class for the modes of using the total 41 features and features selected with SVD are shown in the following tables, respectively. What is evident, this is to increase the precision, recall, and also the F score of intrusion detection system. It is generally seen that, in general, the percentage of diagnosis in which only SVD-selected features are used, more than when we use all 41 features of the data set, particularly, the percentage of attack detection for classrooms U2R and R2L in the use of feature selection is significant and is quite significant. Also, the system uses the preferred features to be able to detect new and unknown attacks that are not encountered during training and there are only experimental data. The following tables show the precision, recall and F- measure for intrusion detection and SVM algorithm in states with and without diminution of dimensions.

Table 2. Precision of attacks detection by SVM by attack class (%)

|  | Normal | DoS | PRB | U2R | R2L |
|---|---|---|---|---|---|
| **total features** | 87.2 | 96.8 | 98.9 | 64.6 | 68.2 |
| **selected features** | 97.4 | 96.9 | 98.3 | 58.3 | 98.1 |

Table 3 Recall of attacks detection by SVM by attack class (%)

|  | Normal | DoS | PRB | U2R | R2L |
|---|---|---|---|---|---|
| **total features** | 97.3 | 88.7 | 98.7 | 42.5 | 55.6 |
| selected features | 98.4 | 86.1 | 99.1 | 46.3 | 69.7 |

Table 4. F- measure of attacks detection by SVM by attack class (%)

|  | Normal | DoS | PRB | U2R | R2L |
|---|---|---|---|---|---|
| total features | 94.5 | 92.1 | 97.1 | 52.2 | 62. 5 |
| selected features | 95.9 | 93.5 | 98.0 | 58.2 | 81.3 |

In this research, the efficiency of SVM algorithm is compared with KNN algorithm. RapidMiner software is used to implement this method. Compare the precision, recall and F-Measure by separating the attack class for a total of 41 features and features selected with SVM and KNN algorithms. Accordingly, the figures are shown below. As can be seen, the percentages of these criteria are more than KNN in almost all features of the SVM.

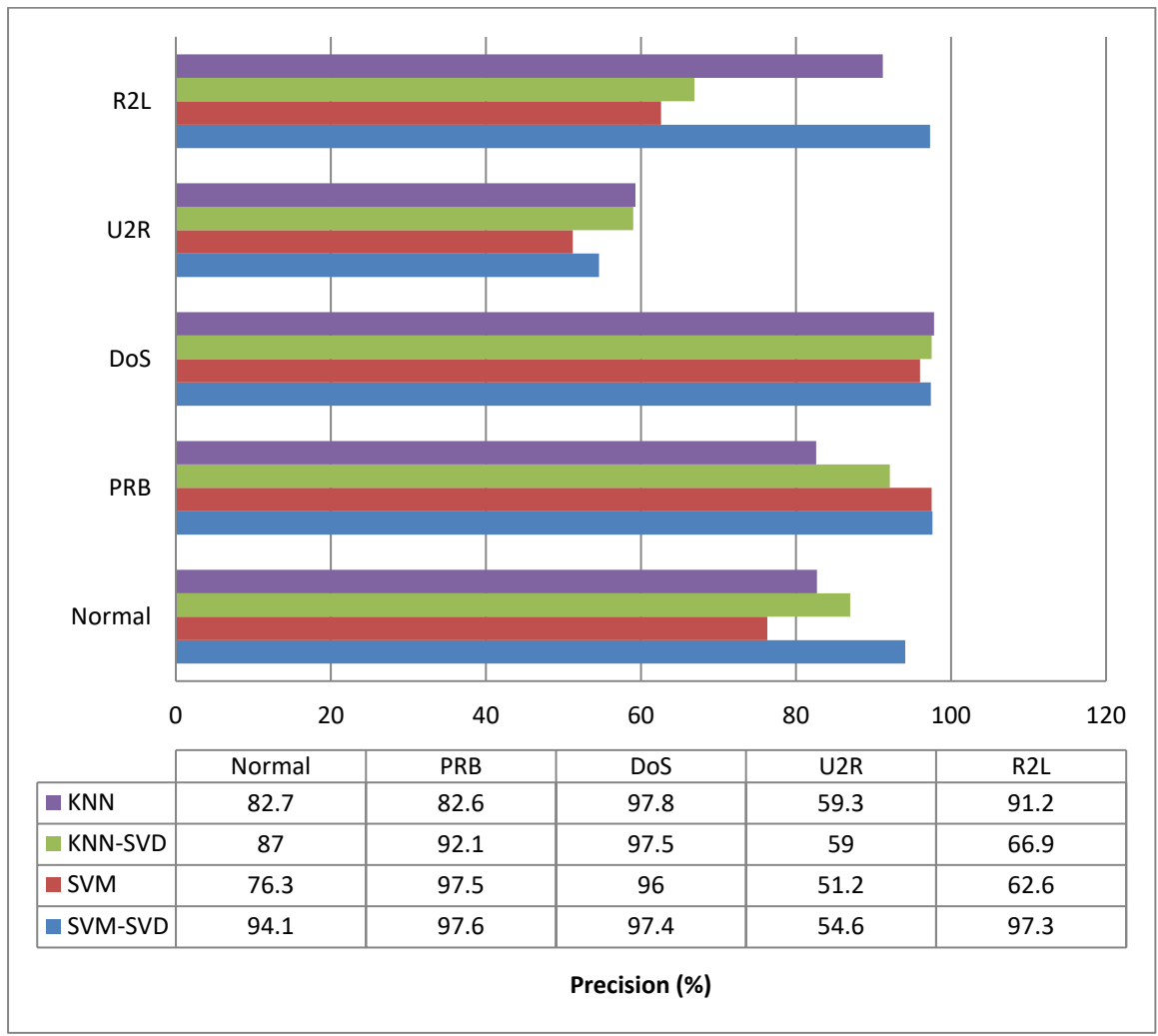

|  | Normal | PRB | DoS | U2R | R2L |
|---|---|---|---|---|---|
| KNN | 82.7 | 82.6 | 97.8 | 59.3 | 91.2 |
| KNN-SVD | 87 | 92.1 | 97.5 | 59 | 66.9 |
| SVM | 76.3 | 97.5 | 96 | 51.2 | 62.6 |
| SVM-SVD | 94.1 | 97.6 | 97.4 | 54.6 | 97.3 |

**Precision (%)**

Figure 6. Comparing the precision measure of attacks detection by attack class for different methods

The graph above shows the ratio of the precision measure to the proposed method compared to the other three methods, and, as it is known, in the average, the proposed method has a better performance.

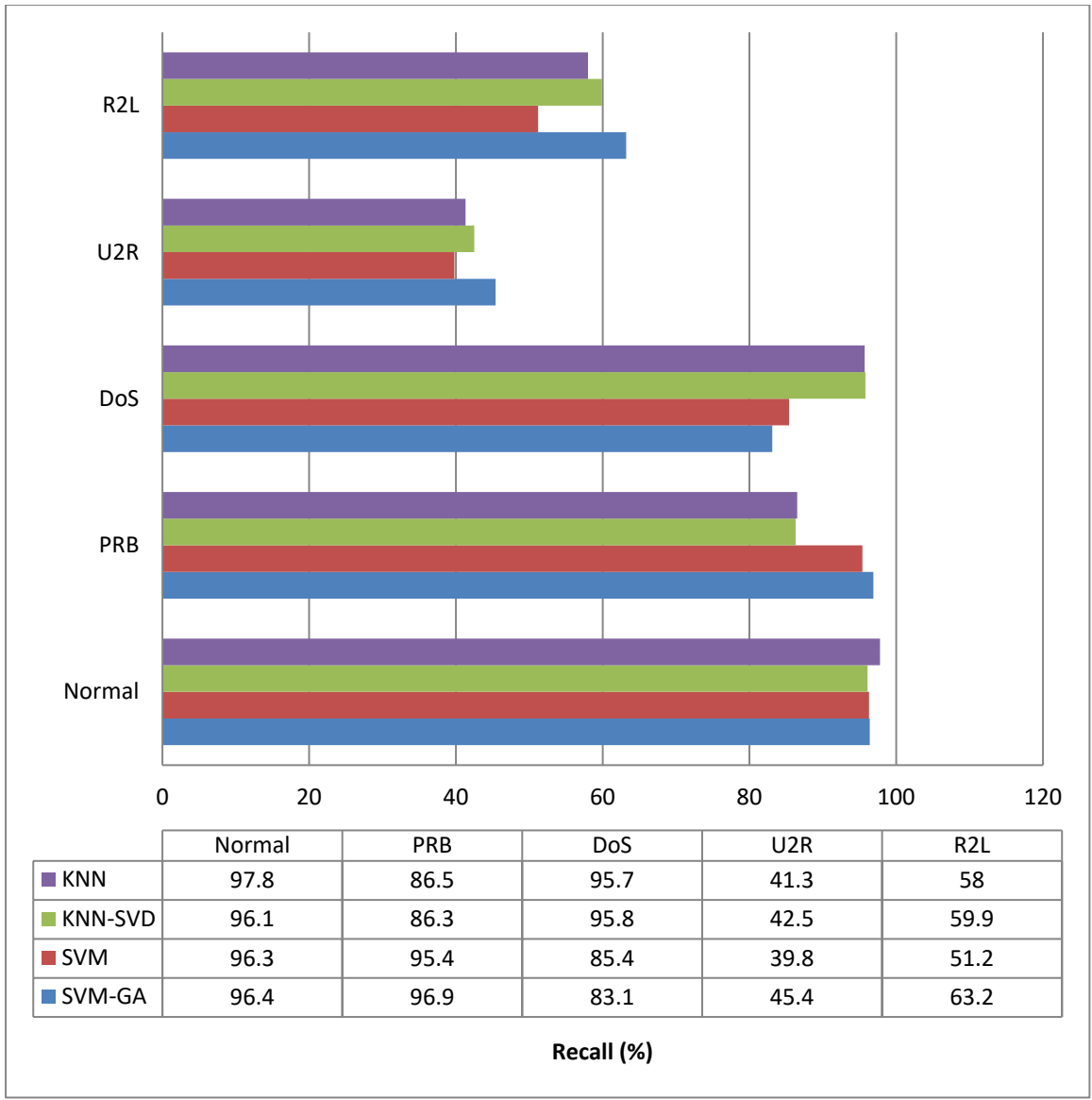

|         | Normal | PRB  | DoS  | U2R  | R2L  |
|---------|--------|------|------|------|------|
| KNN     | 97.8   | 86.5 | 95.7 | 41.3 | 58   |
| KNN-SVD | 96.1   | 86.3 | 95.8 | 42.5 | 59.9 |
| SVM     | 96.3   | 95.4 | 85.4 | 39.8 | 51.2 |
| SVM-GA  | 96.4   | 96.9 | 83.1 | 45.4 | 63.2 |

**Recall (%)**

Figure 7. Comparing the recall measure of attacks detection by attack class for different methods

The graph above shows the ratio of the recall measure to the proposed method compared to the other three methods, and, as it is known, in the average, the proposed method has a better performance.

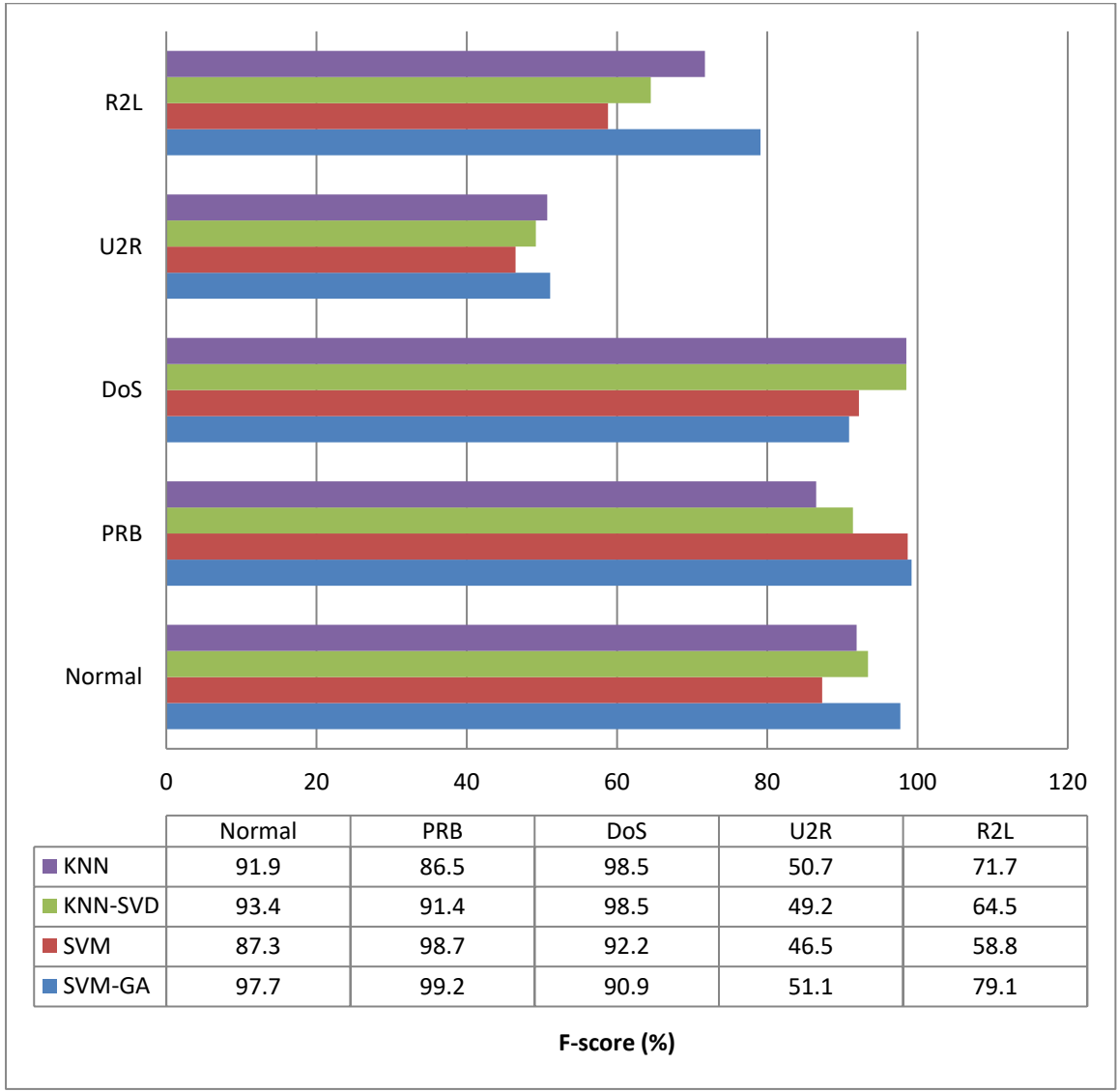

Figure 8. Comparison of the F- measure of attack detection by attack classes for different methods

The above graph shows the ratio of the F- measure to the proposed method compared to the other three methods, and as it is clear, in the average, the proposed method has a better performance.

## 5. Conclusion

The importance of penetration detection systems as an essential and practical tool for ensuring the security of computer networks is undeniable. An intrusion detection system is a hardware or software tool that detects attacks by monitoring the flow of events. Detection of permission allows organizations to maintain their systems against the threats posed by the increasing interconnection between networks and enhance the reliability of their information systems. Intrusion detection systems are security monitoring systems that are used to identify abnormal behaviors and exploit abuse in computers or computer networks. Penetration methods are generally divided into two categories, which include the diagnosis of abuse and the detection of anomalies. Preventive penetration patterns are maintained as a rule, and are applicable to the detection of unauthorized use of the system. So that each pattern can capture different types of permeation and, if such a pattern is observed, the system will be infiltrated. In an abnormal diagnostic method by identifying behavior as a normal behavior for the user, the host, or the whole system, any deviation from this behavior is considered as abnormal behavior, which can be a possible occurrence of an attack. In this research, the system for detecting anomalies using SVD and SVM algorithms was investigated. Different steps of the proposed method, which included preprocessing the data set, selecting the feature, backup vector, and post-processing, were examined separately. The NSL-KDD data set was used to teach and test the proposed model and the results showed that the proposed method has a better performance than the basic methods. The accuracy of the proposed method for the KNN algorithm has improved by an average of 7% for all classes. In this research, SVD algorithm was used to improve the efficiency of the SVM algorithm for intrusion detection. The SVM model is a general way to rank in different research. The SVD algorithm is also not dependent on a particular application can be used in conjunction with SVM (or many other classification methods) to reduce the size and improve the efficiency of classification in other areas of use. Another work that can be considered as future activities is the use of the strengths of several ranking methods as a boosting by combining different classification methods; a better diagnosis can be achieved. Working and testing the proposed method on other types of data sets that have been created more purposefully. It can also be considered as future work. Another future work in this area could be the development of the proposed method for parallel processing and also for use in distributed media such as Hadoop.